\documentclass[doublespacing]{elsart2}
\usepackage{bbm}
\usepackage{amsmath,amssymb,amsfonts,epsf}
\usepackage{graphicx}

\usepackage{amssymb}

\begin{document}

\begin{frontmatter}



\title{Universality in the stock exchange}


\author{}

\address{}

\begin{abstract}
We analyze the constituents stocks of the Dow Jones Industrial
Average (DJIA30) and the Standard \& Poor's 100 index (S\&P100) of
the NYSE stock exchange market. Surprisingly, we discover the data
collapse of the histograms of the DJIA30 price fluctuations and of
the S\&P100 price fluctuations to the universal non-parametric
Bramwell-Holdsworth-Pinton (BHP) distribution. Since the BHP
probability density function appears in several other dissimilar
phenomena, our result reveals an universal feature of the stock
exchange market.
\end{abstract}

\end{frontmatter}

\section{Introduction}
\label{sec:1}

The modeling of the time series of stock  prices is a main issue in
economics and finance and it is of a vital importance in the
management of large portfolio of stocks. From a statistical physics
point of view, one can think that the stock prices form a
non-equilibrium system.
 We investigate the constituent stocks of the
indexes Dow Jones Industrial Average (DJIA30) and Standard \& Poors
100 (S\&P100), of the New York Stock Exchange Market (NYSE),
observed in the 21 years period, January 1987 to September 2008, and
comprising about 5470 trading days. Surprisingly, in this paper, we
observe the data collapse of the histograms of the  DJIA30 price
fluctuations and of the  S\&P100 price fluctuations to the
Bramwell-Holdsworth-Pinton (BHP) probability density function
without any fitting.

\section{Universality of the  Bramwell-Hodsworth-Pinton distribution}

The universal nonparametric BHP pdf was discovered by Bramwell,
Holdsworth and Pinton \cite{BHP1998}. The universal nonparametric
BHP pdf is the pdf of the fluctuations of the total magnetization,
in the strong coupling (low temperature) regime for a
two-dimensional spin model (2dXY), using the spin wave
approximation. The magnetization distribution, that they found, is
named, after them, the Bramwell-Holdsworth-Pinton (BHP)
distribution. The \emph{BHP probability density function (pdf)} is
given by
\begin{eqnarray}
              p(\mu)=\int_{-\infty}^{\infty}\frac{dx}{2\pi}
\sqrt{\frac{1}{2N^2}\sum_{k=1}^{N-1}\frac{1}{\lambda_k^2}}
&\!&e^{ix\mu\sqrt{\frac{1}{2N^2}\sum_{k=1}^{N-1}\frac{1}{\lambda_k^2}}
-\sum_{k=1}^{N-1}\left[\frac{ix}{2N}\frac{1}{\lambda_k}-\frac{i}{2}
\mbox{arctan}\left(\frac{x}{N\lambda_k}\right)\right]}\nonumber\\\!
&\!&.e^{-\sum_{k=1}^{N-1}\left[\frac{1}{4}\mbox{ln}{\left(1+\frac{x^2}{N^2\lambda_k^2}\right)}\right]}\
,
      \label{eq1}
\end{eqnarray}
\noindent where the $\{\lambda_k\}_{k=1}^L$ are the eigenvalues, as
determined in \cite{Bramwelletal2001}, of the adjacency matrix. It
follows, from the formula of the BHP pdf, that the asymptotic values
for large deviations, below and above the mean, are exponential and
double exponential, respectively (in this article, we use the
approximation of the BHP pdf obtained by taking $L=10$ and $N=L^2$
in equation (\ref{eq1})). As we can see, the BHP distribution does
not have any parameter (except the mean that is normalize to 0 and
the standard deviation that is normalized to 1) and it is universal,
in the sense that appears in several physical phenomena. For
instance, the universal nonparametric BHP distribution is a good
model to explain the fluctuations of order parameters in theoretical
examples such as, models of self-organized criticality, equilibrium
critical behavior, percolation phenomena (see \cite{BHP1998}), the
Sneppen model (see \cite{BHP1998} and \cite{DahlstedtJensen2001}),
and auto-ignition fire models (see \cite{SinharayBordaJensen2001}).
The universal nonparametric BHP distribution is, also, an
explanatory model for fluctuations of several phenomenon such as,
width power in steady state systems (see \cite{BHP1998}),
fluctuations in river heights and flow (see \cite{Bramwelletal2001}
and \cite{DahlstedtJensen2005}) and for the plasma density
fluctuations and electrostatic turbulent fluxes measured at the
scrape-off layer of the Alcator C-mod Tokamaks (see
\cite{VanMilligen05}). Recently, Gon\c calves, Pinto and Stollenwerk
\cite{GonPinSto08} observed that the Wolf's sunspot numbers
 fluctuates according to  the universal nonparametric
BHP distribution for, both, the ascending and descending phases.
Surprisingly, we observe  the data collapse of the DJIA30 price
fluctuations and of the S\&P100 price fluctuations to the
Bramwell-Holdsworth-Pinton (BHP) probability density function (pdf).
Hence, our result reveals an universal feature of the daily returns
of the stock prices in the DJIA30 and S\&P100 index.

\section{The S\&P100 stock ensemble}

 The S\&P 100 index is a subset of
the S\&P 500 and is comprised of 100 leading U.S. stocks with
exchange-listed options. The constituents of the S\&P 100 represent
about $57\%$ of the market capitalization of the S\&P 500. The
stocks in the S\&P 100 are generally among the largest companies in
the S\&P 500. The variable investigated in our analysis
 is the \emph{re-scaled daily return} defined by
\begin{equation}
S_i(t)=\Big(\frac{Y_i(t + 1) -Y_i(t)}{Y_i(t)}\Big)^{2/3},
\end{equation}
where the stock $i$ has a closure price $Y_i(t)$ in the day $t$. Let
$n=69$ denote the number of stocks considered in the S\&P 100. We
define the \emph{mean $\mu_{SP}(t)$  of the re-scaled daily return}
by
$$
\mu_{SP}(t)=\frac{1}{n}\sum_{i=1}^nS_i(t)\quad.
$$
\noindent We define the \emph{standard deviation $\sigma_{SP}(t)$ of
the re-scaled daily return}  by
$$
\sigma_{SP}(t)=\sqrt{\frac{\sum_{i=1}^{n}
S_i(t)^2-\mu_{SP}(t)^2}{n}}\quad.
$$
\begin{figure}[!htb]
\begin{center}
\includegraphics[width=12cm]{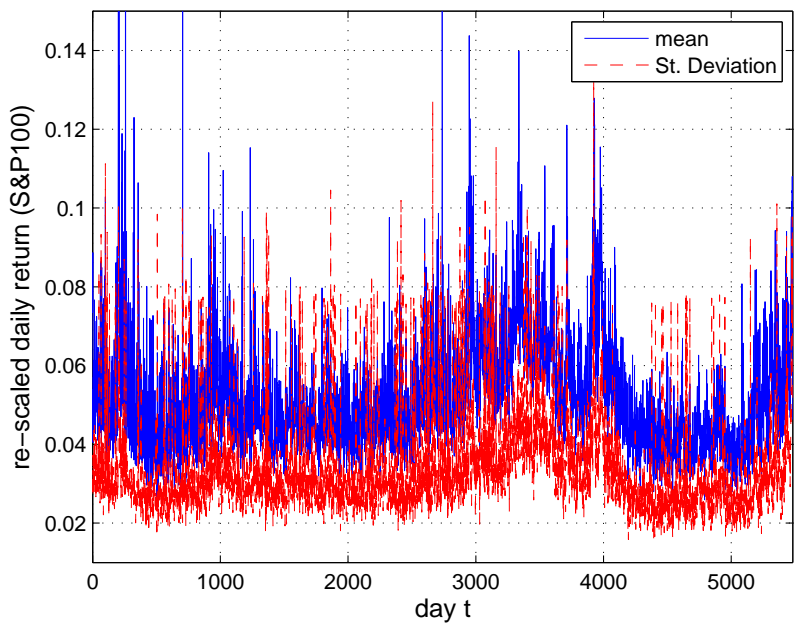}
\caption{\footnotesize{Mean and standard deviation of the re-scaled
daily return of the S\&P100 index.}} \label{fig1}
\end{center}
\end{figure}
\noindent In figure \ref{fig1}, we show the mean and standard
deviation of the re-scaled daily return.
 We define the \emph{S\&P100 fluctuations} $F_i^{SP}(t)$ by
\begin{equation}
          F_i^{SP}(t)=\frac{S_i(t)-\mu_{SP}(t)}{\sigma_{SP}(t)}\quad.
        \label{eq2}
\end{equation}
\noindent

\begin{figure}[!htb]
\begin{center}
\includegraphics[width=12cm]{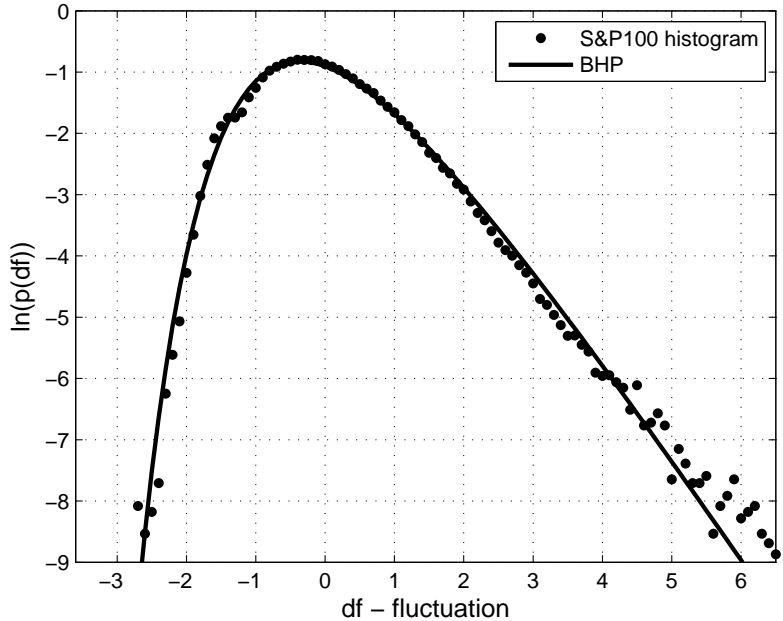}
\caption{\footnotesize{Histogram, in the semi-log scale, of the
S$\&$P100 price fluctuations with the BHP on top.}} \label{fig2}
\end{center}
\end{figure}
\noindent In Figure \ref{fig2}, we show the histogram  of the
S\&P100 price fluctuations with the BHP pdf on top. Since the BHP
pdf is non-parametric, we observe the data collapse of the histogram
of the S\&P100 fluctuations to the BHP pdf (without any fitting).

\section{The DJIA30 stock ensemble}

The DJIA30 consists of 30 of the largest and most widely held public
companies in the United States.
 The variable investigated in our analysis is the \emph{re-scaled daily
return} defined by
\begin{equation}
J_i(t)=\Big(\frac{X_i(t + 1) -X_i(t)}{X_i(t)}\Big)^{2/3},
\end{equation}
where the stock $i$ has a closure price $X_i(t)$ in the day $t$. Let
$n=30$ denote the number of stocks in the DJIA30. We define the
\emph{mean $\mu_{DJ}(t)$ of the re-scaled daily return}  by
$$
\mu_{DJ}(t)=\frac{1}{n}\sum_{i=1}^nJ_i(t)\quad.
$$
\noindent We define the \emph{standard deviation  $\sigma_{DJ}(t)$
of the re-scaled daily return} by
$$
\sigma_{DJ}(t)=\sqrt{\frac{\sum_{i=1}^{n}
J_i(t)^2-\mu_{DJ}(t)^2}{n}}\quad.
$$
\begin{figure}[!htb]
\begin{center}
\includegraphics[width=12cm]{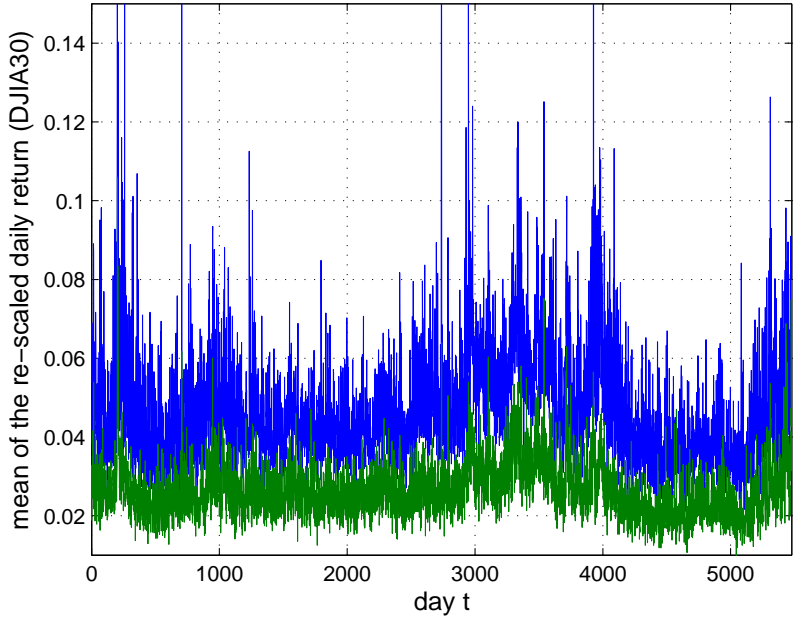}
\caption{\footnotesize{Mean and standard deviation of the re-scaled
daily return of the DJIA30 index.}} \label{fig3}
\end{center}
\end{figure}
\noindent In figure \ref{fig3}, we show the mean and standard
deviation of the re-scaled daily return.
 We define the \emph{DJIA30 price fluctuations} $F_i^{DJ}(t)$ by
\begin{equation}
          F_i^{DJ}(t)=\frac{J_i(t)-\mu_{DJ}(t)}{\sigma_{DJ}(t)}\quad.
        \label{eq2}
\end{equation}
\noindent
\begin{figure}[!htb]
\begin{center}
\includegraphics[width=12cm]{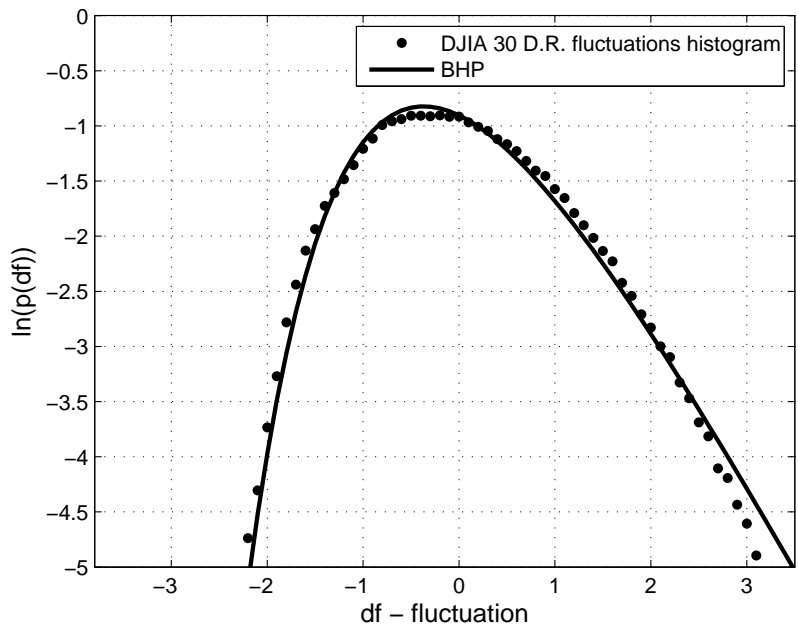}
\caption{\footnotesize{Histogram of the DJIA30 price fluctuations
and of the BHP pdf in the semi-log scale.}}
 \label{fig4}
\end{center}
\end{figure}
\noindent In Figure \ref{fig4}, we show the histogram  of the DJIA30
price fluctuations with the BHP pdf on top. Since the BHP pdf is
non-parametric, we observe the data collapse of the histogram of the
DJIA30 fluctuations to the BHP pdf (without any fitting).

\clearpage

\section{Conclusions}

We analyzed the constituents stocks of the Dow Jones Industrial
Average and the Standard \& Poor's 100 index of the NYSE stock
exchange market. Surprisingly, we discovered the data collapse of
the histograms of the DJIA30 price fluctuations and of the S\&P100
price fluctuations to the universal non-parametric
Bramwell-Holdsworth-Pinton (BHP) distribution.
 Since the BHP probability density function appears
in several other dissimilar phenomena, our result revealed an
universal feature of the stock exchange market.




\end{document}